\documentclass{article}
\usepackage{amsmath,amsfonts}
\usepackage{url}
\usepackage{graphicx}
\usepackage{subfigure}
\usepackage{times}

\usepackage[left=2.5cm,top=2.5cm,right=2.5cm,bottom=2.5cm,nohead,nofoot]{geometry}

\begin{document}
\title{A Prediction Market for Toxic Assets Prices\thanks{This author is an Enterprise Ireland Principal Investigator (grant number PC/2008/0367).}}
\author{Alan Holland\\Cork Constraint Computation Centre,\\ Dept of Computer Science\\University College Cork\\Cork, Ireland\\ \url{a.holland@4c.ucc.ie}\\ }

%

\maketitle              

\begin{abstract}
We propose the development of a prediction market for forecasting
prices for ``toxic assets'' to be transferred from Irish banks to
the National Asset Management Agency (NAMA). Such a market allows
market participants to assume a stake in a security whose value is
tied to a future event. We propose that securities are created whose
value hinges on the transfer amount paid for loans from NAMA to a
bank. In essence, bets are accepted on whether the price is higher
or lower than a certain quoted figure. The prices of the securities
represent transfer prices for toxic assets increases or decreases in
line with market opinion. Prediction markets offer a proven means of
aggregating distributed knowledge pertaining to fair market values
in a scalable and transparent manner. They are incentive compatible
(i.e. induce truthful reporting) and robust to strategic
manipulation. We propose that a prediction market is run in parallel
with the pricing procedure recommended by the European Commission.
This procedure need not necessarily take heed of the prediction
markets view in all cases but it may offer guidance and a means of
anomaly detection. An online prediction market would offer everybody
an opportunity to ``have their say'' in an open and transparent
manner.
\end{abstract}

\section{Introduction}
The Irish Government has recently decided to set up a National Asset
Management Agency whose objective is to acquire ``toxic assets''
from Irish banks on a mandatory basis so that they are removed from
the balance sheet of at an agreed price. These assets are impaired
loans that were given to developers so they could purchase land or
fund construction projects. There are four key challenges associated
with this approach.

\begin{enumerate}
  \item It is necessary to determine a fair price for these assets
in the absence of a liquid property market and price signals in a
clear and transparent manner.
  \item Determining the optimal
basket of toxic assets to acquire given a finite budget, estimated
fair prices and a multi-criteria objective function that seeks to
maximize returns to the tax-payer, minimize asset management
duration and allow recapitalization of the banks on the open market.
  \item The problem of establishing a
scalable solution that can expedite the asset transfer process.
Given the large number of properties that are dispersed across
Ireland and many other countries that require valuation, it is
imperative that any suggested approach can be rapidly expanded to
deal with the enormous volume of troubled loans.
  \item The pricing and optimisation mechanisms need to be transparent so that it is clear
to all tax-payers that no preferential treatment is being offered to
any party and the process does not suffer from political
interference.
\end{enumerate}

In order to address these challenges, we suggest that a prediction
market be initiated so that distributed knowledge among property
experts (and even non-experts) can be capitalised on in a manner
that incentives truthful reporting of valuations. Invited
participants in the market will be offered an opportunity to wager
on the transfer prices of toxic assets. This offers a means of
leveraging ``the wisdom of crowds'' and improving the accuracy of
fair price predictions in the absence of a liquid property market.
Such an automated mechanism for price prediction offer scalability
and a clear incentive for NAMA to determine prices that are aligned
with public/market opinion. This addresses many of the criticisms
leveled at NAMA by various
commentators~\cite{gurdiev09nama,whelan09pt}.

The prediction market could be operated on a public-access basis or
available only to registered experts in property valuation. We
assume in this work that it is open to anybody (over the age of 18)
who wishes to participate. Given the ease with which scalability can
be provided, it is difficult to justify any argument in favour of
excluding the general public. It is more politically acceptable and
also improves the accuracy and robustness of the results.

We also briefly describe an extension that could enhance this
approach. It involves determining dependencies between assets and
finding super- or sub-additive valuations. We briefly outline a
method that seeks to optimise the value of a portfolio
Section~\ref{sec:conc} concludes.

\section{Key Benefits of Prediction Markets}
The quest to predict the future is pervades all aspects of society.
According to Bragues, ``\emph{A potential solution to this
epistemological conundrum has emerged through mass
collaboration}''~\cite{bragues09pm}. Prediction markets allow
participants to purchase a stake in a security whose value is tied
to a future event. The fluctuating prices offer a continuously
updated probability estimate of the likely outcome of the event. A
key advantage of prediction markets over other approaches to
information aggregation (e.g. surveys) is that they provide
incentives for \emph{truthful revelation }of
beliefs~\cite{wolfers08pm}. If prediction markets are used as inputs
into future decisions, this may provide a countervailing incentive
to trade dishonestly to manipulate market prices. However, Hanson
and Oprea demonstrated that such \emph{attempts at manipulation are
destined to fail} because other participants have a greater
incentive to counteract such efforts~\cite{hanson05manip}. An
important role for prediction markets is that potential profits
offer an incentive for \emph{information discovery}. The aggregated
finding of all such efforts at information discovery would provide a
strong indication to the NAMA as to the markets expectation
regarding the fair price that will be paid. Some of the other
benefits are listed below.
\begin{enumerate}
  \item Responsiveness: it has been shown that prediction markets
  respond quickly to new information~\cite{snowberg06resp}.
  \item Arbitrage opportunities tend to be fleeting thus indicating
  market efficiency.
  \item Improved accuracy: the incentive compatibility removes some
  of the behavioural anomalies that are witnessed in surveys.
  Prediction markets for events such as political elections are
  widely regarded as more accurate than survey techniques.
  \item Versatility: they can be adopted in a wide variety of
  settings. For example, Hewlett Packard yielded more accurate sales
  forecasts than the firms internal experts.
\end{enumerate}

\section{Implementation Considerations}
We present a prototype prediction market and we invite the reader to
visit \url{https://nama.inklingmarkets.com} to see how this market
operates for just a single property.

\subsection{Prototype}
\begin{figure}
\centering
  \includegraphics[width=12cm]{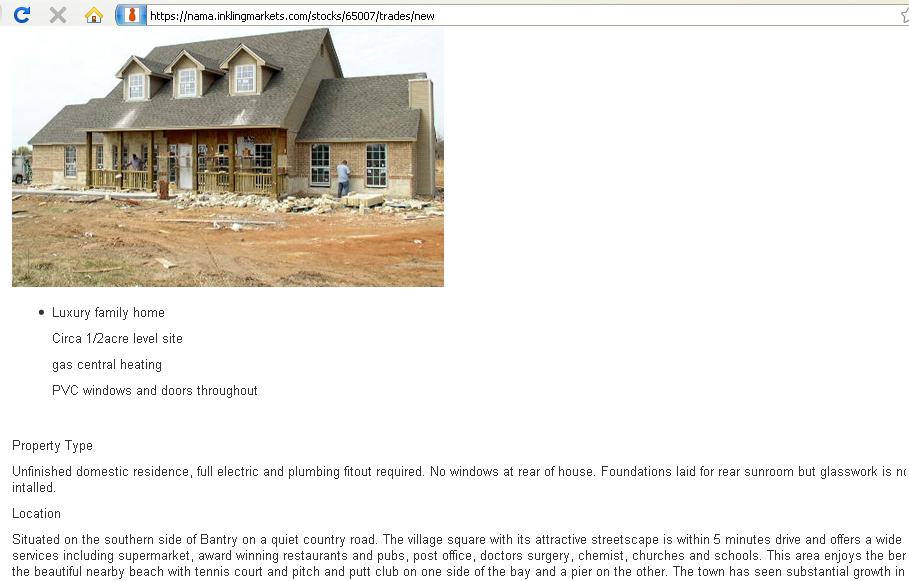}\\
  \caption{Fictional Unfinished Property}\label{fig:bantry_prop}
\end{figure}

The author has prepared a prototype model of a prediction
marketplace\footnote{See \url{https://nama.inklingmarkets.com/}.}
using a tool developed by Inkling Inc. It involves a single
fictional unfinished property in Bantry, Co. Cork
(Fig~\ref{fig:bantry_prop}).
\begin{figure}
\centering
  \includegraphics[width=12cm]{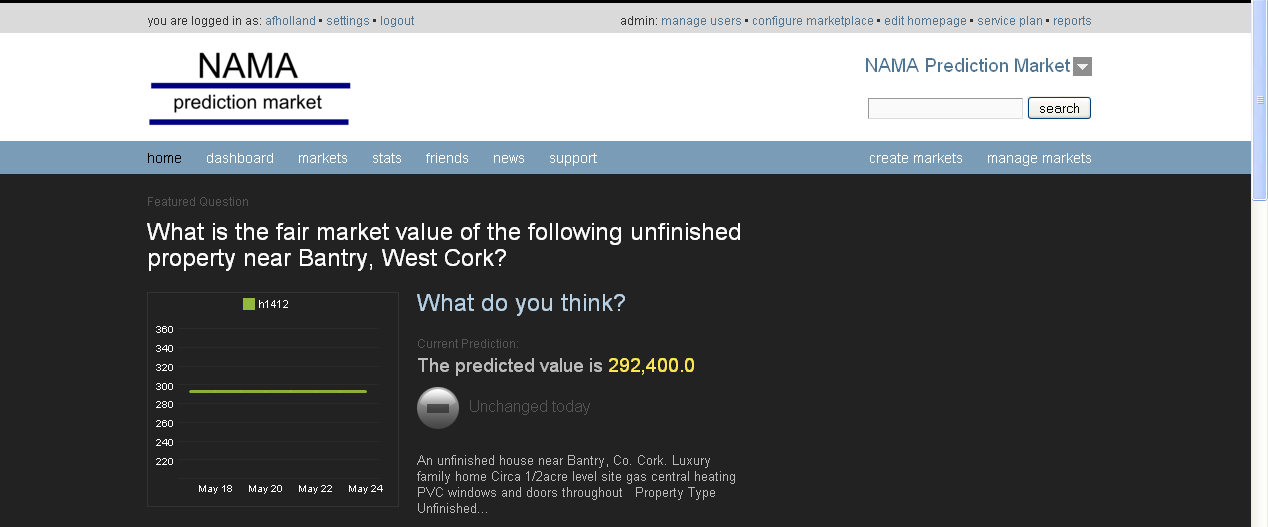}\\
  \caption{NAMA Prediction Market Prototype}\label{fig:nama_home}
\end{figure}

\begin{figure}
\centering
  \includegraphics[width=12cm]{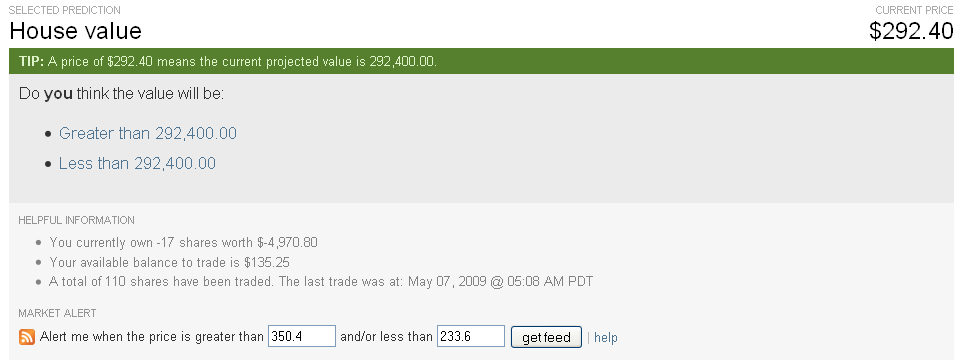}\\
  \caption{Higher/Lower Choice}\label{fig:highLow}
\end{figure}

\begin{figure}
\centering
  \includegraphics[width=12cm]{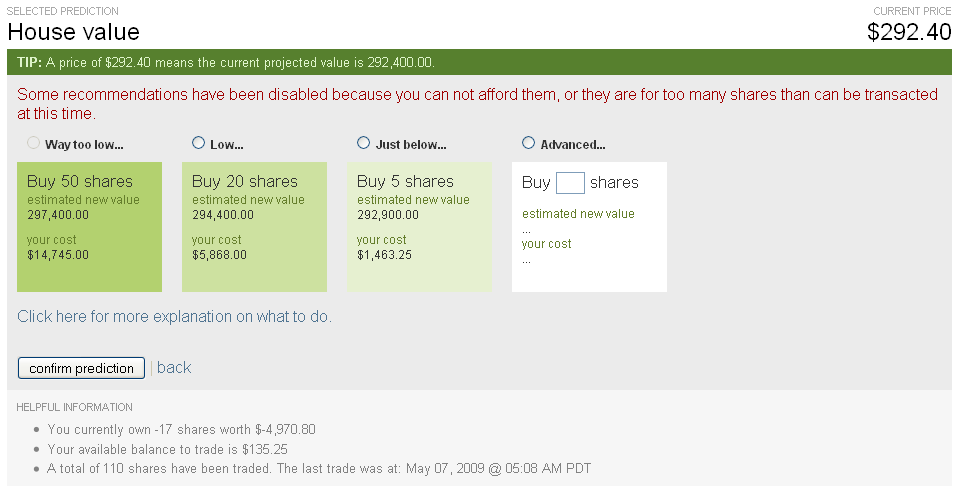}\\
  \caption{Investment Amount}\label{fig:amt}
\end{figure}

Market participants can choose whether to wager on the agreed price
being higher or lower than the current threshold
(Fig.~\ref{fig:nama_home}). They can also choose how much to invest
in that decision, Figures~\ref{fig:highLow} and~\ref{fig:amt}. We
would envisage that each participant is limited in what they can
wager on any single market. To encourage liquidity, it may be
necessary to offer starting balances that can be topped up so that
users have an incentive to engage initially and learn how the market
works. From that point onwards, participants would have an incentive
to enter markets pertaining to securities about which they were most
knowledgeable. This would probably involve studying assets in their
vicinity. In this manner, knowledge aggregation on a grand scale
would be automated and trustworthy.

\subsection{Software System}
It is important that a software system that would host such a market
is secure, scales well and provides a rich set of features. For this
reason it may be preferable to utilise existing systems from
experienced providers. Two obvious possibilities include the
following.

\begin{description}
  \item[Inkling Markets: ] Inkling provides a hosted software solution that could host such a
prediction market(\url{http://inklingmarkets.com/}). Their customers
include Cisco, CNN and Procter \& Gamble.
  \item[Intrade:] Alternatively, Intrade Ltd is one of the world's largest prediction
market makers and is based in Dublin\footnote{The author has had no
contacts with either company and is not acting on their behalf.}
This may offer an excellent opportunity in terms of access to the
necessary software engineering expertise for rapid deployment and/or
customisation of a bespoke solution.
\end{description}

A desirable bespoke element of a tailored solution would include
geographical browsing of assets that are close to a participants
location. Other desiderata include a complete specification of the
European Commission guidelines for estimating asset values. This
would aid participants understanding of the goals of the NAMA in
price determination. Of course it may be possible for the NAMA to
condition it's final value on the prediction markets view. This may
serve to slow convergence to an outcome but this is a decision that
should be taken in light of the liquidity of that market.

The immediate scalability depends largely upon the format and
content of data regarding the underlying assets that were acquired
by property developers. If all data is held in electronic form and
conforms to a set of schema detailing the semantics of this
information then transferal to an online database is relatively
straightforward. If, however, the relevant data is unstructured or
not held electronically then the data aggregation phase would be
slower. Temporal aspects are also important for the correct
functioning of the market. It is necessary for market participants
to know a cut-off date by which it will become public knowledge how
much was paid for certain assets.

The prediction market relies on open and transparent access to data
so the legitimacy of information disclosure and potential privacy
breaches would need further study.

\section{Combinatorial Prediction Markets and Portfolio Optimisation}
It is possible to set up more complex contracts that depend on the
outcome of more than one event. This offers insight into the
dependencies or correlation between events. For example, when
valuing properties there are often super- or sub-additive valuations
ascribed to collections of properties. The value may be enhanced if
the assets complement one another. For example, adjacent properties
may offer access to a road or other resources that increase the
overall value of the combination. In the case where assets are
substitutes, it may be desirable to acquire just one of them.

In order to optimise values, it is necessary to elicit information
regarding dependencies. The necessary queries (i.e. combinatorial
markets) will need to be decided using heuristics based on proximity
or adjacency. The problem of optimising combined asset value minus
the sum of the values of the individual assets can be modelled as a
Mixed Integer Program. The computational problem is complex but
efficient solvers exist for problems with tens of thousands of
variables. Ultimately, this approach would need to be investigated
but is of secondary importance compared to the valuation of
stand-alone properties.

\section{Conclusion}\label{sec:conc}
This proposal highlights the possibility of a complementary online
market that offers indicative prices for toxic assets. A prediction
market would improve the \emph{accuracy}, \emph{speed} and
\emph{scalability} of pricing toxic assets. It would also offer
\emph{transparency} and a voice to the general public who wish to
have an input in this matter. There is little to lose but much to
gain in adopting this approach that can run in parallel with
existing pricing mechanisms in a seamless manner.

\bibliographystyle{plain}
\bibliography{nama}
\end{document}